\documentclass[journal,twocolumn]{IEEEtran}
\usepackage{graphicx,latexsym,subfigure,amsmath,epsfig,latexsym,subfigure,cite,amssymb}
\usepackage{CJK}
\usepackage{url}
\usepackage{float}
\usepackage{mathrsfs}

\usepackage{graphicx}
\newtheorem{theorem}{Theorem}

\newtheorem{proof}{Proof}
\newtheorem{corollary}{Corollary}
\newtheorem{lemma}{Lemma}
\usepackage{cases}
\usepackage{setspace}
\usepackage{subfigure}
\usepackage{diagbox}
\usepackage{color}
\usepackage{caption}
\usepackage[ruled,linesnumbered]{algorithm2e}
\usepackage{bm}
\usepackage{caption}
\usepackage{amsmath}
\allowdisplaybreaks[4]
\ifCLASSINFOpdf
\else
\fi

\begin{document}

\title{Reinforcement Learning-based Energy Trading for Microgrids}
\author{Liang Xiao\IEEEauthorrefmark{1}, Xingyu Xiao\IEEEauthorrefmark{1}, Canhuang Dai\IEEEauthorrefmark{1},
Mugen Peng\IEEEauthorrefmark{2}, {Lichun Wang}\IEEEauthorrefmark{3} and H. Vincent Poor\IEEEauthorrefmark{4},\\

\IEEEauthorblockA{\IEEEauthorrefmark{1} Dept. Communication Engineering, Xiamen University, China. Email: lxiao@xmu.edu.cn}\\
\IEEEauthorblockA{\IEEEauthorrefmark{2} Dept. of Information and Communication Engineering, Beijing  University of Posts and Telecommunications, China. Email: pmg@bupt.edu.cn}\\
\IEEEauthorblockA{\IEEEauthorrefmark{3} Dept. of Electrical and Computer Engineering,
National Chiao Tung University, Hsinchu, Taiwan. Email: lichun@cc.nctu.tw}\\
\IEEEauthorblockA{\IEEEauthorrefmark{4} Dept. of Electrical Engineering, Princeton
University, Princeton, NJ. Email: poor@princeton.edu}
}


\maketitle
\begin{abstract}
With the time-varying renewable energy generation and power demand, microgrids (MGs) exchange energy in smart grids to  reduce their dependence on power plants. In this paper, we formulate an MG energy trading game, in which each MG trades energy  according to the predicted renewable energy generation and local energy demand,  the current battery level, and the  energy trading history. The Nash equilibrium (NE) of the game is provided, revealing the conditions under which   the local energy generation satisfies the  energy demand of the MG and providing  the performance bound of the energy trading scheme. We propose a reinforcement learning based MG energy trading scheme that applies the
deep Q-network (DQN)  to  improve the utility of the MG for the case with a large number of the connected MGs. Simulations are performed for the MGs with wind generation that are aware of  the electricity prices and the historic energy trading, showing that this scheme significantly reduces the average power plant schedules and improves the utility of the MG compared with the  benchmark strategy.
\end{abstract}


\IEEEpeerreviewmaketitle

\section{Introduction}
\IEEEPARstart{M}{icrogrids} (MGs) as small-scale power supply networks consist of renewable energy generators (e.g., wind turbines and solar panels), local electrical consumers (e.g., air conditioners), and energy storage devices (e.g., batteries) \cite{amin2005toward}. Each MG can be aware of the local energy supply,  the demand profile and the energy trading prices with the other MGs and  the connecting power plant\cite{farhangi2010path}. With  time-varying renewable energy production and demand, an MG can  sell its extra energy to the other MGs to reduce the dependence on the power plant and save the long-distant energy transmission loss \cite{hatziargyriou2007microgrids}.

Each MG  can choose the intended amount of the trading energy with the connected MGs and the power plant based on its current battery level, the expected renewable energy generation such as the model proposed in \cite{wang2016incentivizing}, and the energy trading history, and then negotiate with the other MGs to determine the actual amount of the trading energy.  In this paper, we formulate the interactions among the connected MGs and the power plant  as an energy trading game.
In this game, an MG has to make a tradeoff between the trading profit and the energy gain, following the local power demand.
We provide the Nash equilibrium (NE) of the  game to disclose how the electricity price, the renewable power production, the local electricity demand, and the battery level impact on the energy trading, and the conditions under which the NE exists to show how an MG can rely on the local energy trading to satisfy the energy demand.

Reinforcement learning (RL) techniques such as Q-learning have been used  to optimize the MG energy storage and generation  \cite{qiu2016heterogeneous,
dalal2016hierarchical,kim2016dynamic,xu2012multiagent,guan2015reinforcement}.
The MG energy trading decision in the repeated game can be formulated as a Markov decision process (MDP), in which an MG is unaware of the battery levels, the energy generation and the energy demands of the other MGs. Therefore,
an MG can apply  Q-learning to choose the amount of energy bought or sold with the other MGs and the power plant without being aware of the energy generation and demand models of the other MGs.
In this scheme, the decision is made according to  the current state  that consists of the predicted renewable power generation and  energy demand, the current battery level, and the quality function or Q function for each state-strategy pair. Updated according to the iterative Bellman equation, the Q function provides the expected discounted long-term reward of an MG from an energy trading decision in a time slot.

To accelerate the learning speed of the Q-learning based scheme for the case that involves a large number of MGs and feasible trading strategies, we propose a deep Q-network (DQN) based MG energy trading scheme. As a deep reinforcement learning \cite{mnih2015human},  DQN extracts features from the high-dimension state-action space in smart grids.
This scheme exploits a deep convolutional neural network (CNN) to estimate the Q-value for each trading policy.
Simulations are performed for the smart grid that consists of the MGs equipped with wind turbines, in which the  wind speed and the electricity prices are retrieved from Hong Kong Observation and ISO New England, respectively, over time, for a given  MG battery capacity.
Simulation results show that the DQN-based scheme reduces the power plant schedules, and increases the average utility of MGs, compared with the Q-learning based scheme proposed in \cite{xiao2017energy}.

The main contributions of this paper are listed as follows:
{\begin{itemize}
\item We formulate an MG energy trading game  with the MG energy trading decision made based on battery level, local demand, renewable energy generation model and energy trading history. We provide the conditions under which an NE exists, showing how an MG can satisfy its local demand by the renewable energy generation of the local smart grid.
\item We propose a DQN-based MG energy trading scheme in the dynamic game to reduce the dependence on  power plants and increase the utility of the MG compared with the benchmark algorithm in \cite{xiao2017energy}.
\end{itemize}}

The rest of this paper is organized as follows: We review the related work in Section \ref{related}, and present the system model in Section \ref{SYS}. We formulate an MG energy trading game in Section \ref{NEs} and propose a DQN-based energy trading scheme  in Section \ref{DQNsys}. We provide simulation results in Section \ref{simulation} and conclude this work in Section \ref{conclusion}.

\section{Related Work}\label{related}

Energy trading has attracted significant research attention recently \cite{zachar2017microgrid,wang2016integrated,werth2015conceptual,kuznetsova2014integrated},
and game theory is  a powerful tool to study the energy exchange in \cite{wang2016incentivizing,chakraborty2015real,Zhang2014Randomized,wang2014game,
xiao2015prospect,xiao2014anti,tushar2015three,baeyens2011wind,Zhang2016A,
Park2016Contribution,maharjan2013dependable}.
For example,  the MG energy exchange is formulated in  \cite{wang2016incentivizing} as a Nash bargaining game, in which the microgrids cooperatively decide the trading amounts and the prices in the smart grid.
The MG energy trading game based on  coalitional game theory in \cite{chakraborty2015real}  aims to reduce the long-distant transmission loss.
The auction mechanism can help motivate microgrids to participate in the energy trading, including the randomized auction based scheme as designed in \cite{Zhang2014Randomized} and the double-auction based market as developed in \cite{wang2014game}.
The prospect theory based MG energy exchange game as formulated in \cite{xiao2015prospect} analyzes the subjectivity decision of end-users in the energy exchange under uncertain energy production and demand with dynamic prices.
The energy cheating attack  in the energy exchange game can be suppressed by the reputation mechanism and the indirect reciprocity principle \cite{xiao2014anti}.

Reinforcement learning techniques have been used  to optimize the energy storage and generation \cite{guan2015reinforcement,qiu2016heterogeneous,kim2016dynamic,dalal2016hierarchical,
xu2012multiagent}. For example, the Q-learning  based storage control scheme has been applied in \cite{qiu2016heterogeneous} to improve the system efficiency of a heterogeneous smart grid with multiple battery types. The RL-based energy generation scheme in \cite{dalal2016hierarchical} considers a two-layer Markov model and chooses whether to participate in the next day power generation to improve the day-ahead and real-time reliability. The Q-learning based  pricing scheme as presented in \cite{kim2016dynamic} can encourage customers to use electricity more efficiently.

A hotbooting Q-learning based energy trade as presented in \cite{xiao2017energy}
improves the utility of the MG and reduces the long-distant power transmission loss in the dynamic energy trading game.
Compared with our previous work in \cite{xiao2017energy}, we formulate a  stochastic MG energy trading game by considering the estimation error of the renewable energy generation and propose a DQN-based energy trading strategy  to  improve the trading performance for the case that involves a large number of MGs. Simulations are performed for smart grids according to the energy generation and demand data collected from practical energy systems.

\section{System Model}\label{SYS}
\begin{figure}\setlength{\abovecaptionskip}{0cm}\setlength{\belowcaptionskip}{-0.6cm}
\begin{center}
\includegraphics[width=3.1 in]{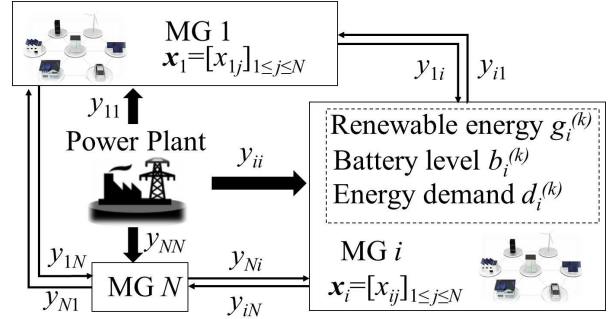}\\
\caption{Energy trading  in a smart grid consisting of  $N$ connected MGs and a power plant, where MG $i$ that determines its energy trading intention $\boldsymbol{x}_i$ buys or sells $|y_{ij}|$ amounts of energy with MG $j$ or the power plant. }\label{model}
\end{center}
\end{figure}
We consider  \emph{N} MGs that are connected with each other and a power plant. Each MG is equipped with renewable power generators, active loads, electricity storage devices, and the power transmission lines connecting with the other  MGs and the power plant, as shown in Fig. \ref{model}.
The time horizon of each day is divided into $T$ equal time slots, with each time slot lasts $24/T$ hours.
A microgrid can receive energy  from the other microgirds, the power plant, and the local  energy generators that exploit wind, photovoltaic, biomass, and tidal energy, and can observe the battery level, the energy trading prices, and its current energy production and  energy demand in the energy trading game.

The renewable energy generator such as the wind turbine supplies  local-independent, intermittent and time-varying energy. The amount of the renewable energy generated by MG $i$ at time $k$ denoted by $g_{i}^{(k)}$ can be estimated via the power generation history and the modeling method such as  \cite{wang2016incentivizing}. The estimated amount of the generated power is denoted by $\hat{g}_{i}^{(k)}$.

The amount of energy that MG $i$ intends to sell to (or buy from) MG $j$ before the bargaining is denoted by $x_{ij}^{(k)}$, and that to trade with the power plant is denoted by $x_{ii}^{(k)}$.  If $x_{ij}^{(k)}<0$, MG $i$ intends to sell its extra energy to MG $j$ or the power plant. If $x_{ij}^{(k)}>0$, MG $i$ aims to buy energy.
The trading strategy of MG $i$ at time $k$ is denoted by $\boldsymbol{x}_{i}^{(k)}=[x_{ij}^{(k)}]_{1\leq j\leq N}\in \boldsymbol{X}$, where $\boldsymbol{X}$ is the MG action set.

As two MGs usually contradicts in the energy trading decisions, e.g., $x_{ij}^{(k)}x_{ji}^{(k)}>0$, energy trading bargaining has to be made and the resulting MG trading strategy  is denoted by $ \boldsymbol{y}_{i}^{(k)}=[y_{ij}^{(k)}]_{1\leq j\leq N}$, where $y_{ij}^{(k)}$ (or  $y_{ii}^{(k)}$) denotes the amount of the energy actually trades with
MG $j$ (or the power plant). MG $i$ purchases energy if $y_{ij}^{(k)}>0$ and sells energy if $y_{ij}^{(k)}<0$. We assume that $|y_{ij}^{(k)}|\leq C$, where $C$ is the maximum amount of the MG energy exchange between two MGs. Time index $k$ is omitted, if no confusion incurs.
The actual energy trading between MG $i$ and MG $j$ after the bargaining depends on their trading intentions as follows,
 \begin{equation}\label{action}
 y_{ij}=\begin{cases} \max(x_{ij}, -x_{ji}),\,\, \mbox{if }x_{ij}<0,\,x_{ji}>0,\,\forall\, i\neq j \\ \min(x_{ij},-x_{ji}),\,\, \mbox{if }x_{ij}>0,\,x_{ji}<0, \,\forall\, i\neq j \\\sum_{1\leq j\leq N}x_{ij}-\sum_{1\leq i\neq j\leq N}y_{ij}, \,\, \mbox{if }\forall\, i=j
 \\0, \qquad\qquad\qquad\qquad\mbox{o.w.}\end{cases}
\end{equation}
It is clear that  $y_{ij}+y_{ji}=0$, $\forall\, i\neq j$.

Each MG has energy storage devices such as batteries to charge energy under low load and discharge under high load. The battery level of MG $i$, denoted by $b_{i}^{(k)}$ is limited by the storage capacity denoted by $B$, with $0< b^{(k)}_{i}\leq B$. According to the  demand record at the same time in history,  MG $i$ estimates the amount of the energy demand at time $k$ denoted by $\hat{d_{i}}^{(k)}$, with $0\leq \hat{d_{i}}^{(k)} \leq D_{i}$, where $D_{i}$ is the peak energy demand of  MG $i$.
The actual amount of the energy demand is denoted by $d_{i}^{(k)}$.
Based on the energy trading, the local energy generation, and the energy  demand, the battery level of MG $i$ in a  smart grid with $N$ MGs is given by
\begin{equation}\label{battery}
b^{(k+1)}_{i}=b^{(k)}_{i}+g^{(k)}_{i} - d^{(k)}_{i}+\sum_{j=1}^{N}y_{ij}^{(k)}.
\end{equation}

The energy gain of MG $i$, denoted by $G_i(b)$, is defined as the benefit that the MG obtains from  battery level $b$. It is obvious that the energy gain is nondecreasing with $b$, and $G(0)=0$. Note that the logarithmic function is widely used in economics to model the preference ordering of users and decision making process \cite{maharjan2013dependable}. Therefore, we assume that
\begin{align}
G_i(b)=\beta_i\ln(1+b),
\end{align}
where the positive coefficient $\beta_i$ represents the importance for MG $i$  to satisfy the energy demand of the consumers.

To encourage the energy exchange among the MGs, the local market provides a higher selling price for
the MG trading denoted by $\rho^{(k)}_{-}$ and a lower buying price denoted by $\rho^{(k)}_{+}$, compared with the prices offered by the power plant that are denoted by $\xi^{(k)}_{-}$ and $\xi^{(k)}_{+}$, respectively, i.e., $\rho^{(k)}_{-}>\xi^{(k)}_{-}$ and $\rho^{(k)}_{+}<\xi^{(k)}_{+}$. For simplicity, we define the energy price vector $\boldsymbol{\rho}^{(k)}=[\rho^{(k)}_{-}, \rho^{(k)}_{+}, \xi^{(k)}_{-}, \xi^{(k)}_{+}]$.

\section{Stochastic Energy Trading Game}\label{NEs}
The MG energy trading in a smart grid with $N$ connected MGs and a power plant can be formulated as an energy trading game consisting of $N$ players. Each MG chooses the amount of energy to sell to or purchase from the other MGs and the power plant, $\boldsymbol{x}_{i} \in \boldsymbol{X} $ based on the current battery level, the predicted energy generation model and the local  power demand.
The actual trading energy scheme of MG $i$, ($\boldsymbol{y}_i$) results from the negotiation based on the energy trading intention of all the $N$ MGs.

The utility of MG $i$ at time $k$, denoted by $u_{i}^{(k)}$, depends on the energy gain and the trading profit.
Let $\boldsymbol{y}$ be the actual energy trading scheme of $N$ MGs after negotiation and $\textrm{I}(\cdot)$ be an indicator function that equals 1 if the argument is true and 0 otherwise. We assume that
\begin{align}\label{reward}
u_{i}^{(k)}(\boldsymbol{y})=&\beta \ln\left(1+b_{i}^{(k)}+g_{i}^{(k)}-d_{i}^{(k)}+\sum_{j=1}^{N}y_{j}\right)\cr
&-\sum_{j\neq i}^{N}y_{j}\left(\textrm{I}(y_{j}\leq 0)\rho^{(k)}_{-}+\textrm{I}(y_{j}> 0)\rho^{(k)}_{+}\right)\cr
&-y_{i}\left(\textrm{I}(y_{i}\leq 0)\xi^{(k)}_{-}+\textrm{I}(y_{i}> 0)\xi^{(k)}_{+}\right).
\end{align}
\setlength{\parskip}{0.5\baselineskip}
The first term in the right-hand-side (RHS) of (\ref{reward}) is the energy gain, the second term is the trading profit or payment from other MGs, and the third term corresponds to that with the power plant.

In this work, we assume equal energy selling and buying prices among the MGs, i.e., $\rho_{-}= \rho_{+}=\rho$, and that the selling price is lower than the buying price with the power plant are  $\xi_{-}=\rho(1-\varepsilon)$ and $\xi_{+}=\rho(1+\varepsilon)$, where $\varepsilon$ is the buying/selling price ration, with $0<\varepsilon<1$. In this case, we have
\begin{align}\label{price}
\boldsymbol{\rho}=[ \rho,\, \rho,\, \rho(1-\varepsilon),\, \rho(1+\varepsilon) ].
\end{align}

\setlength{\parskip}{0.5\baselineskip}
Denote the NE of the energy trading game with $N=3$ MGs by $[\boldsymbol{x}^*_{i}]_{1\leq i\leq N}=[ x_{ij}^{*}]_{1\leq j \leq N}$. By definition, each MG chooses its energy trading strategy at the NE state to maximize its own utility, if the other MGs apply the NE strategy.
Therefore,  for any $\boldsymbol{x}_{i} \in \boldsymbol{X}$, we have
\begin{align}
\label{u1*}u_{1}(\boldsymbol{x_{1}}^{*}, \boldsymbol{x_{2}}^{*},  \boldsymbol{x_{3}}^{*})\geq u_{1}(\boldsymbol{x_{1}}, \boldsymbol{x_{2}}^{*}, \boldsymbol{x_{3}}^{*})\\
\label{u2*}u_{2}(\boldsymbol{x_{1}}^{*}, \boldsymbol{x_{2}}^{*}, \boldsymbol{x_{3}}^*)\geq u_{2}(\boldsymbol{x_{1}^*}, \boldsymbol{x_{2}}, \boldsymbol{x_{3}}^*)\\
\label{u3*}u_{3}(\boldsymbol{x_{1}}^*, \boldsymbol{x_{2}}^*, \boldsymbol{x_{3}}^{*})\geq u_{3}(\boldsymbol{x_{1}}^*, \boldsymbol{x_{2}}^{*}, \boldsymbol{x_{3}})
\end{align}

We first evaluate the NE of the static energy trading game with $N=3$ microgrids that are connected to a power plant.
\begin{theorem}\label{th1}
The energy trading game with $N=3$  has an NE given by
\begin{align}
\label{x1*}\boldsymbol{x_{1}}^{*}=&\bigg[\frac{\beta(3-2\varepsilon)}{\rho(1-\varepsilon)}-3
-\sum_{i=1}^{3}\Big(b_{i}^{(k)}+g_{i}^{(k)}-d_{i}^{(k)}\Big),\cr
&\quad\frac{\beta}{\rho}-1- b_{1}^{(k)}-g_{1}^{(k)}+d_{1}^{(k)},\cr
&\quad\frac{\beta}{\rho}-1- b_{1}^{(k)}-g_{1}^{(k)}+d_{1}^{(k)}\bigg]\\
\label{x2*}\boldsymbol{x_{2}}^{*}=&\bigg[\frac{\beta}{\rho}-1- b_{2}^{(k)}-g_{2}^{(k)}+d_{2}^{(k)},0,0\bigg]\\
\label{x3*}\boldsymbol{x_{3}}^{*}=&\bigg[\frac{\beta}{\rho}-1- b_{3}^{(k)}-g_{3}^{(k)}+d_{3}^{(k)},0,0\bigg]
\end{align}
if
\begin{align}
\label{con1}&\frac{1-\varepsilon}{3-2\varepsilon}\bigg(3+\sum_{i=1}^{3}\Big(b_{i}^{(k)}+g_{i}^{(k)}-d_{i}^{(k)}\Big)\bigg)>
\frac{\beta}{\rho}>\cr
&\qquad1+\max_{i=2,3}\Big(b_{i}^{(k)}+g_{i}^{(k)}-d_{i}^{(k)}\Big).
\end{align}
\begin{proof}
If (\ref{con1}) holds, we have
\begin{align}\label{simp1}
&u_{1}^{(k)}\left(\boldsymbol{x}_{1}, \boldsymbol{x}_{2},  \boldsymbol{x}_{3}\right)=\beta\ln\Big(1+b_{1}^{(k)}+g_{1}^{(k)}-d_{1}^{(k)}+x_{11}\cr
& + \max (x_{12},-x_{21})+\max(x_{13},-x_{31})\Big)-\rho(1-\varepsilon) x_{11}\cr
&-\rho\big(\max (x_{12},-x_{21})+\max(x_{13},-x_{31})\big).
\end{align}
If (\ref{x2*}) and (\ref{x3*}) hold, we have
\begin{align}
&u_{1}^{(k)}\big(\boldsymbol{x}_{1}, \boldsymbol{x}^*_{2},  \boldsymbol{x}^*_{3}\big)
=\beta\ln\Big(1+b_{1}^{(k)}+g_{1}^{(k)}-d_{1}^{(k)}+x_{11}\cr
&+x^*_{21}+x^*_{31}\Big)-\rho(1-\varepsilon) x_{11}+\rho(x^*_{21}+x^*_{31}),
\end{align}
and then
\begin{align}
\frac{\partial^{2} u^{(k)}_1\big(\boldsymbol{x}_{1}, \boldsymbol{x}_{2}^*,\boldsymbol{x}_{3}^*\big)}{\partial \bm{x}_{1}^{2}}\succeq 0,
\end{align}
indicating that as $\partial u^{(k)}_1\left(\boldsymbol{x}_{1}, \boldsymbol{x}_{2}^*,\boldsymbol{x}_{3}^*\right)/\partial \bm{x}_{1} = 0$,
\begin{align}
u_{1}^{(k)}\big(\boldsymbol{x}_{1}, \boldsymbol{x}^*_{2},  \boldsymbol{x}^*_{3}\big)\leq u_{1}^{(k)}\big(\boldsymbol{x}_{1}^*, \boldsymbol{x}^*_{2},  \boldsymbol{x}^*_{3}\big).
\end{align}
Thus, (\ref{x1*}) holds for (\ref{u1*}). Similarly, we can prove that (\ref{u2*}) and (\ref{u3*}) hold indicating that (\ref{x1*})-(\ref{x3*}) is the NE of the game.
\end{proof}
\end{theorem}

\begin{corollary}
If (\ref{con1}) holds, the actual amounts of the energy that MG 1 sells to the power plant and the other 2 MGs  are
$\beta(3-2\varepsilon)/\rho(1-\varepsilon)-3-\sum_{i=1}^{3}(b_{i}^{(k)}+g_{i}^{(k)}-d_{i}^{(k)})$,
$-\beta/\rho+1+b_{2}^{(k)}+g_{2}^{(k)}-d_{2}^{(k)}$, and $-\beta/\rho+1+b_{3}^{(k)}+g_{3}^{(k)}-d_{3}^{(k)}$, respectively.
\end{corollary}
\begin{figure}[htbp]\setlength{\abovecaptionskip}{0cm}\setlength{\belowcaptionskip}{-0cm}
\begin{center}
\includegraphics[width=3.5 in]{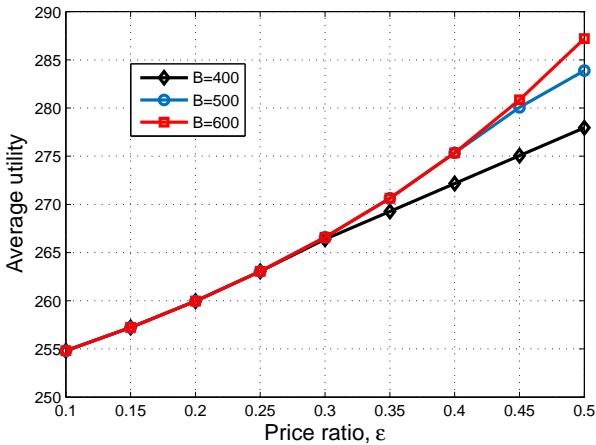}\\
\caption{Average utility of an MG in  the energy trading game consisting of  three MGs, with $\beta=120$.}\label{NE}
\end{center}
\end{figure}
As shown in Fig. \ref{NE}, the average utility of an MG increases with the battery capacity and the power selling/buying price ratio, because the MG can store more electricity in the off-peak time to improve the  profit in the  trading.

We now consider a stochastic game denoted by $\mathbb{G}$, in which the MGs estimate  the amount of the renewable energy generated in the time slot and the maximum estimation error denoted by $\triangle^{(k)}$ follows the distribution denoted by $f$ given by
\begin{align}\label{g}
f(g)=
\begin{cases}
P,  &g=\hat{g}\\
(1-P)/2,  & g= \hat{g}-\triangle\quad \text{or} \quad\hat{g}+\triangle \\
0,  & \text{o.w.}
\end{cases}.
\end{align}
In other words, the probability of an accurate estimation of the renewable energy generation is $P$, and the MG has an estimation error $\pm \triangle$ with probability $(1-P)/2$, where $P$ denotes the estimation accuracy.

\begin{lemma} The stochastic energy trading game $\mathbb{G}$ has an NE given by
\begin{align}
&\boldsymbol{x}_{1}^{*}=\Bigg[
\frac{3P(1-\varepsilon)}{2\beta(1-P)}-\frac{1}{1-P}\Big(P^2-P+\frac{\rho^2(1-\varepsilon)^2}{4\beta^2}\Big)^{1/2}\cr
&-3-\sum_{1\leq i\leq 3}\bigg(b_{i}^{(k)}+g_{i}^{(k)}-d_{i}^{(k)}\bigg)
,\,d_{1}^{(k)}-b_{1}^{(k)}\cr
&-1-g_{1}^{(k)}+\frac{1}{1-P}\bigg(\Big(P^2-P+\frac{\rho^2}{4\beta}\Big)^{1/2}
+\frac{\rho}{2\beta}\bigg),\cr
&\frac{1}{P-1}\Big(\big(P^2-P+\frac{\rho^2}{4\beta}\big)^{1/2}-\frac{\rho}{2\beta}\Big)-1
-b_{1}^{(k)}\cr
&-g_{1}^{(k)}+d_{1}^{(k)}\Bigg]
\end{align}
\begin{align}
\boldsymbol{x}_{i}^{*}=&\Bigg[ \frac{1}{1-P}\Bigg(\bigg(P^2-P+\frac{\rho^2}{4\beta^2}\bigg)^{1/2}+\frac{\rho}{2\beta}\Bigg)\cr
&-1-b_{i}^{(k)}-g_{i}^{(k)}+d_{i}^{(k)},\, 0,\, 0\Bigg],\,i=2,3,
\end{align}
if
\begin{align}
&\frac{3P(1-\varepsilon)}{2\beta}-\bigg(P^2-P+\frac{\rho^2(1-\varepsilon)^2}{4\beta^2}\bigg)^{1/2}\cr
&\leq (1-P)\bigg(3+\sum_{1\leq i\leq 3}\Big(b_{i}^{(k)}+g_{i}^{(k)}-d_{i}^{(k)}\Big)\bigg)\\
&\bigg(P^2-P+\frac{\rho^2}{4\beta^2}\bigg)^{1/2}+\frac{\rho}{2\beta}
>\cr
&(1-P)\bigg(1+\max_{i=2,3}\Big(b_{i}^{(k)}+g_{i}^{(k)}-d_{i}^{(k)}\Big)\bigg).
\end{align}
\begin{proof}
Similar with that to Theorem 1.
\end{proof}
\end{lemma}

\section{DQN-based energy trading Scheme}\label{DQNsys}
\begin{figure*}[htb]
  \centering
  \includegraphics[width=5.0 in]{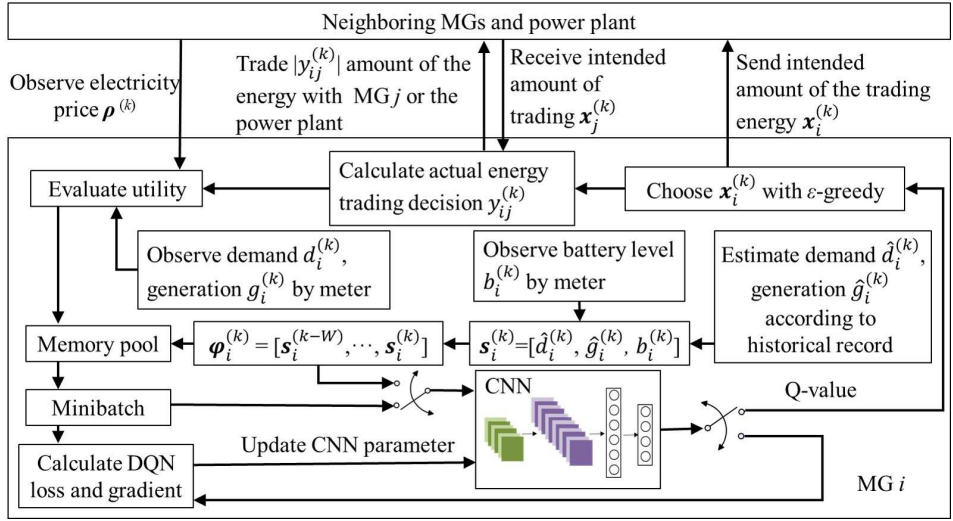}\\
  \caption{DQN-based MG energy trading scheme.}\label{DQN}
\end{figure*}
\begin{algorithm}[!h]
\caption {DQN-based energy trading strategy}
\label{DQNal}
Initialize $\alpha$, $\gamma$,\, $W$, $\theta^{(1)}$, and $\mathcal{D}={\O}$\\
\For{$k=1, 2, \cdots $}
{
Estimate energy generation $\hat{g}_{i}^{(k)}$ and demand  $\hat{d}_{i}^{(k)}$\\
Observe battery level $b_{i}^{(k)}$\\
$\boldsymbol{s}_{i}^{(k)}=\Big[\,\hat{d}_i^{(k)},\,\hat{g}_i^{(k)},\,b_i^{(k)}\Big]$\\
\uIf{$k\leq W$}
{
Choose $\boldsymbol{x}_{i}^{(k)} \in \boldsymbol{X}^{N}$ at random
}
\Else
{
Form experience sequence $\boldsymbol{\varphi}_{i}^{(k)} = \Big(\boldsymbol{s}_{i}^{(k-W)},\boldsymbol{s}_{i}^{(k-W+1)},\cdots, \boldsymbol{s}_{i}^{(k)}\Big)$\\
Input $\boldsymbol{\varphi}_{i}^{(k)}$ to the CNN with weights $\boldsymbol{\theta}^{(k)}$\\
Obtain the CNN outputs of $\boldsymbol{\varphi}_{i}^{(k)}$, $Q\left(\bm{\varphi}_i^{(k)}, \bm{x}; \boldsymbol{\theta}_{i}^{(k)}\right)$ \\
Choose $\boldsymbol{x}_{i}^{(k)}$ via (\ref{epsilon-greedy})\\
}
Send the intended amount of the energy trading $\boldsymbol{x}_{i}^{(k)}$\\
\For {$j=1, 2, \cdots, N $}
{
Receive the intended amount of the energy trading from the neighboring MGs $x_{ji}^{(k)}$\\
Calculate $y_{ij}^{(k)}$ via (\ref{action})\\
\uIf{$j\neq i$}
{
\uIf {$y_{ij}^{(k)}<0$}
{Sell $-y_{ij}^{(k)}$ amount of the energy to MG $j$}
\Else
{Purchase $y_{ij}^{(k)}$ amount of the energy from MG $j$}}
\Else
{
\uIf {$y_{ii}^{(k)}<0$}
{Sell $-y_{ii}^{(k)}$ amount of the energy to the power plant}
\Else
{Purchase $y_{ii}^{(k)}$ amount of the energy from the power plant}
}
}
Observe  actual generation $g_{i}^{(k)}$, actual demand $d_{i}^{(k)}$\\
Observe the electricity price $\boldsymbol{\rho}^{(k)}$ \\
Evaluate $u_i^{(k)}$ via (\ref{reward})\\
$\mathcal{D}\leftarrow \mathcal{D}\cup \boldsymbol{e}^{(k)}$\\
Select $M$ experience sequences from $\mathcal{D}$ at random forming $\mathcal{M}$\\
Calculate $L\Big(\boldsymbol{\theta}^{(k)}\Big)$ via (\ref{loss})\\
Update the CNN weights with $\boldsymbol{\theta}^{(k)}$ by minibatch gradient descent\\
}
\end{algorithm}
The MG trading decision in a dynamic energy trading game $\mathbb{G}$ can be formulated as a Markov decision process, in which MG $i$ decides the intended amount of the trading energy denoted by $\boldsymbol{x}^{(k)}_{i}=[x_{ij}^{(k)}]_{1\leq j\leq N }$, and exchanges such information with the  neighboring $N-1$ MGs.
Therefore, we propose a DQN-based energy trading scheme that uses CNN as a nonlinear function approximator to estimate the quality or Q value of each feasible energy trading policy and compress the state space of the MG.


In each  time slot,   according to the generation model and historical record, MG $i$  can estimate the local energy demand $\hat{d}_{i}^{(k)}$ and the  renewable energy generation $\hat{g}_{i}^{(k)}$, and observe the current battery level $b_{i}^{(k)}$, which are used to formulate the state denoted by $\boldsymbol{s}^{(k)}_{i}$, with $\boldsymbol{s}^{(k)}_{i}=\Big[ \hat{d}^{(k)}_{i}, \hat{g}^{(k)}_{i}, b^{(k)}_{i}\Big]$.
The experience sequence at time $k$, denoted by  $\bm{\varphi}_i^{(k)}$  consists of the current state $\boldsymbol{s}^{(k)}_{i}$, and the previous $W$ state-action pairs, with $\bm{\varphi}_i^{(k)}=\Big(\boldsymbol{s}_i^{(k-W)},\boldsymbol{x}_i^{(k-W)},
\cdots,\boldsymbol{x}_i^{(k-1)},\boldsymbol{s}_i^{(k)}\Big)$.

MG $i$ chooses the energy trading strategy $\bm{x}_{i}^{(k)}$ in first $W$ time slots at random, and reshapes the state sequence $\bm{\varphi}_i^{(k)}$ afterwards into a square matrix. The matrix is  the input of the CNN with weights denoted by $\bm{\theta}_{i}^{(k)}$. As shown in Fig. \ref{DQN}, the CNN consists of two convolutional (Conv) layers and two fully connected (FC) layers. Similar to \cite{mnih2015human},
the first Conv layer includes $n^{[1]}$ filters, each with size $f^{[1]} \times f^{[1]}$ and stride $s^{[1]}$, and
uses a rectified linear units (ReLU) as an activation function.
The second Conv layer involves $n^{[2]}$ filters, each with $f^{[2]} \times f^{[2]}$ and stride $s^{[2]}$, again followed by a ReLU. The outputs of Conv 2 are input to two FC layers, which use $n^{[3]}$ and $|X|^{N}$ rectified linear units respectively, where $|X|^{N}$ is the number of feasible trading actions for MG $i$.
The output of  the CNN denoted by $Q\left(\bm{\varphi}_i^{(k)}, \bm{x}; \boldsymbol{\theta}_{i}^{(k)}\right)$ is the estimated Q-value for the $|X|^N$ trading actions.

According to the $\epsilon$-greedy algorithm by which  the scheme can avoid staying in the local optimum, the MG $i $ chooses the ``optimal" strategy
as the intended amount of the
trading energy, $\bm{x}^{(k)}_i = [x_{ij}^{(k)}]_{1\leq i\leq N}$, that
maximizes the  Q-value  with a high probability $1 - \epsilon$, and another strategy  with a low probability $\epsilon /(|X|^{N}-1)$, i. e.,
\begin{small}
\begin{align}\label{epsilon-greedy}
\mathrm{Pr}\Big(\bm{x}^{(k)}\Big)=
\begin{cases}
1-\epsilon,  &\bm{x}^{(k)}=\arg\max\limits_{\boldsymbol{x}'\in \bm{X}} Q\Big(\bm{\varphi}_{i}^{(k)},\bm{x}'\Big)\\
\frac{\epsilon}{|X|^{N}-1},  &\text{o.w.}
\end{cases}
\end{align}
\end{small}

The MGs exchange their decision $\boldsymbol{x}_{j}^{(k)}$  with $1\leq j \leq N$ to calculate the actual amount for the energy trading  $\boldsymbol{y}_{i}^{(k)}$ via  Eq. (\ref{action}). If $y_{ij}>0$ (or $y_{ii}>0$), MG $i$ purchases $y_{ij}$ amount of energy from MG $j$ (or the power plant), otherwise, sells $-y_{ij}$ amount of energy.

After the energy trade, MG $i$ observes the actual energy generation $g_{i}^{(k)}$, local demand $d_{i}^{(k)}$ and the energy trading price $\bm{\rho}^{(k)}$ to evaluate the utility $u_{i}^{(k)}$ via Eq. (\ref{reward}).

The DQN-based energy trading scheme stores such energy trading experience of MG $i$  denoted by $\boldsymbol{e}_i^{(k)}=\Big(\boldsymbol{\varphi}_i^{(k)}, \boldsymbol{x}_i^{(k)}, u_{i}^{(k)}, \boldsymbol{\varphi}_i^{(k+1)}\Big)$ in the replay memory pool denoted by $\mathcal{D}=\Big\{\boldsymbol{e}_i^{(1)},\cdots , \boldsymbol{e}_i^{(k)}\Big\}$.
According to the experience replay, MG $i$ randomly chooses $M$ experience sequences from $\mathcal{D}$, denoted by $\mathcal{M}=\big\{\boldsymbol{e}^{(m)}\big\}_{m \in \{1,\cdots, M\}}$, where $M$ is the size of the minibatch.
The loss function on the minibatch, i.e., the mean-squared error of the target optimal Q-value, denoted by $L\Big(\boldsymbol{\theta}_{i}^{(k)}\Big)$ is given by
\begin{align}\label{loss}
L\Big(\boldsymbol{\theta}_{i}^{(k)}\Big)=
&\mathbb{E}_{\mathcal{M}}\Bigg[\bigg(u_{i}+\gamma \max_{\boldsymbol{x}'\in \bm{X}}Q\Big(\boldsymbol{\varphi}'_{i}, \boldsymbol{x}'; \boldsymbol{\theta}_{i}^{(k-1)}\Big)\cr
&\qquad\qquad-Q\Big(\boldsymbol{\varphi}_{i}, \boldsymbol{x}; \boldsymbol{\theta}_{i}^{(k)}\Big)\bigg)^2\Bigg],
\end{align}
where $\boldsymbol{\varphi}'_{i}$ is the next state sequence, and the discount factor $\gamma\in[0,1]$ represents the uncertainty of the future utility.
According to the minibatch gradient descent  algorithm given by \cite{mnih2015human},
the CNN weights $\theta^{(k)}$ is updated by minimizing the
loss function $L\Big(\bm{\theta}^{(k)}
_i\Big)$, as summarized in  Algorithm \ref{DQNal}.

\section{Simulation Results}\label{simulation}

\begin{figure}[htb]
    \begin{minipage}[t]{0.45\linewidth}
  \centering
  \subfigure[Wind power generation]{
    \label{wind} 
    \includegraphics[width=1.7in]{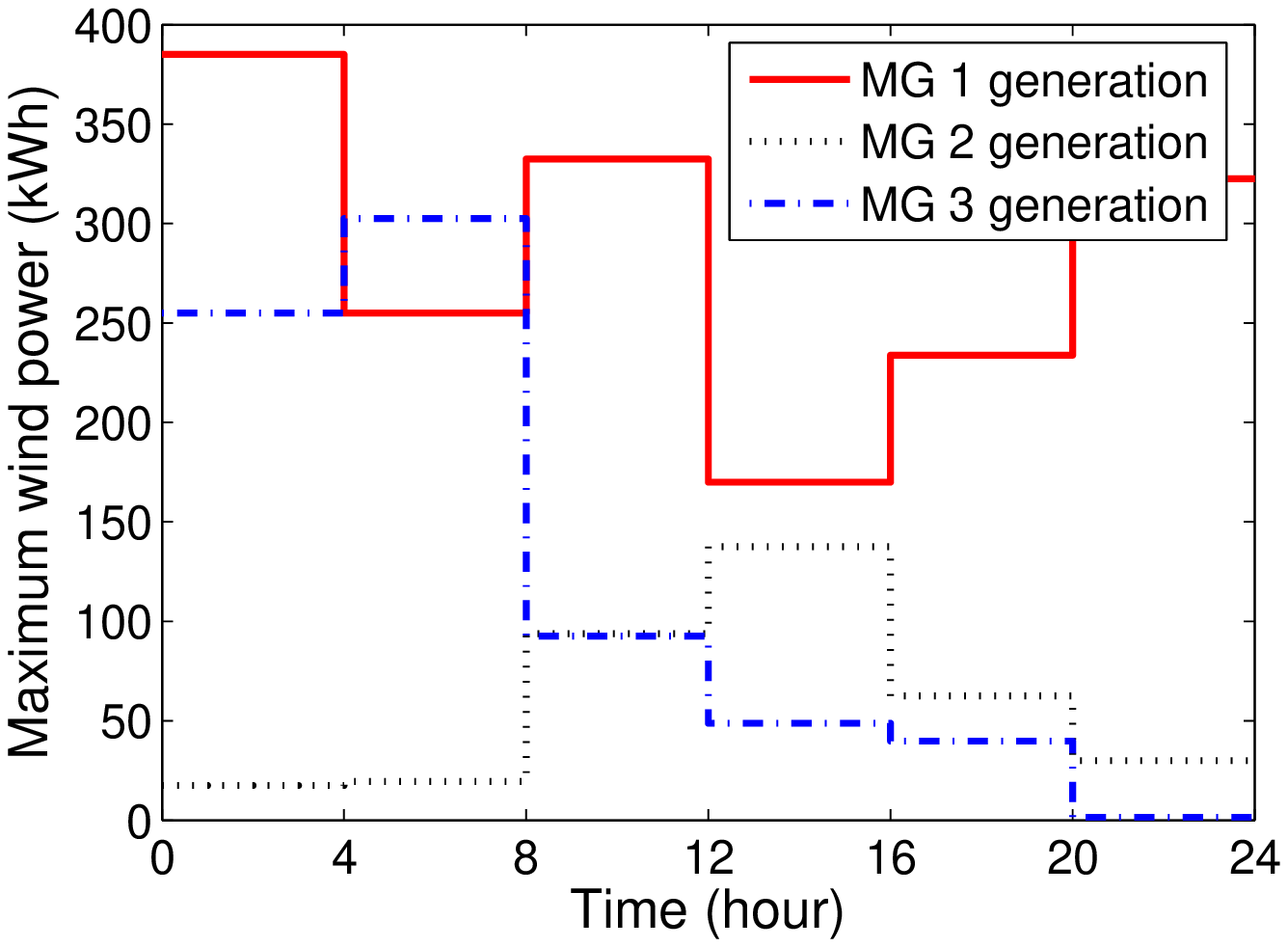}}
        \end{minipage}
    \begin{minipage}[t]{0.45\linewidth}
  \subfigure[Local demand]{
    \label{local} 
    \includegraphics[width=1.7in]{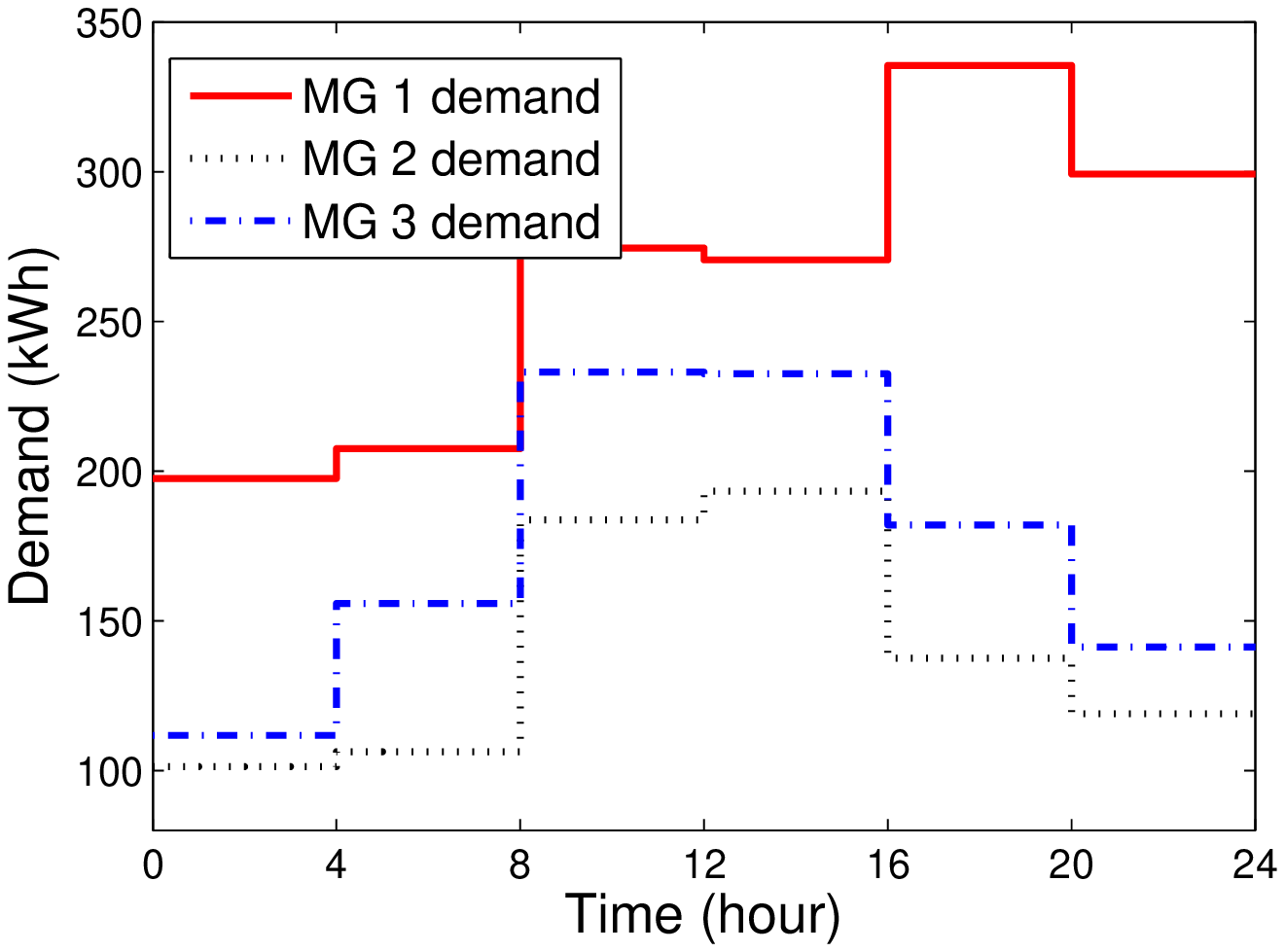}}
        \end{minipage}
  \caption{Supply and demand profiles of MG 1, MG 2 and MG 3.}
  \label{data} 
\end{figure}

We evaluate the performance of the energy trading strategy in the smart grid consisting of three interconnected MGs and a power plant.
The simulations are based on the historical wind speed data collected by Hong Kong Observation, as shown in Fig. \ref{wind}, in which each MG applies the wind power generation model in \cite{Wang2015Joint} to predict the renewable power generation in each hour.
In the simulations, the  daily power demand of each MG is modeled based on the real historical records published in \cite{wang2016incentivizing}, as shown in Fig. \ref{local}, and the MG electricity trading prices follow the ISO New England record in  \cite{ISO-url},
with the MG energy trading price changing over time between 0.19 to 0.44 HKD/kWh.
The parameters of the DQN-based energy trading strategy are listed in Table \ref{CNNtb} with $\beta=120$ and $T=6$.
The hotbooting Q-learning based trading scheme in \cite{xiao2017energy} are evaluated in the simulations as a benchmark.

\begin{table}[htb]\label{cnn}
\small
\renewcommand{\arraystretch}{1.1}
\centering
\begin{tabular}{|c|c|c|c|c|}
\hline
\textbf{Layer} & \textbf{Conv 1} & \textbf{Conv 2} & \textbf{FC 1} & \textbf{FC 2}  \\
\hline
\textbf{Input} & $6\times6$ & $4\times4\times20$ & 360 & 180 \\
\hline
\textbf{Filter size} & $3\times3$ & $2\times2$ & \ /  &  \ / \\
\hline
\textbf{Stride} & $1$ & $1$ & \ / &  \ / \\
\hline
\# \textbf{filters} & $20$ & $40$ & $180$ & $|X|^{N}$  \\
\hline
\textbf{Activation} & ReLU & ReLU & ReLU & \ /  \\
\hline
\textbf{Output}  & $4\times4\times20$ & $3\times3\times40$ & $180$ & $|X|^{N}$  \\
\hline
\end{tabular}
\caption{CNN ARCHITECTURE PARAMETERS}
\label{CNNtb}
\end{table}

\begin{figure*}[htbp]
    \begin{minipage}[t]{0.3\linewidth}
  \centering
    \subfigure[Time in a day]{
    \label{tradepp} 
    \includegraphics[width=2.3in]{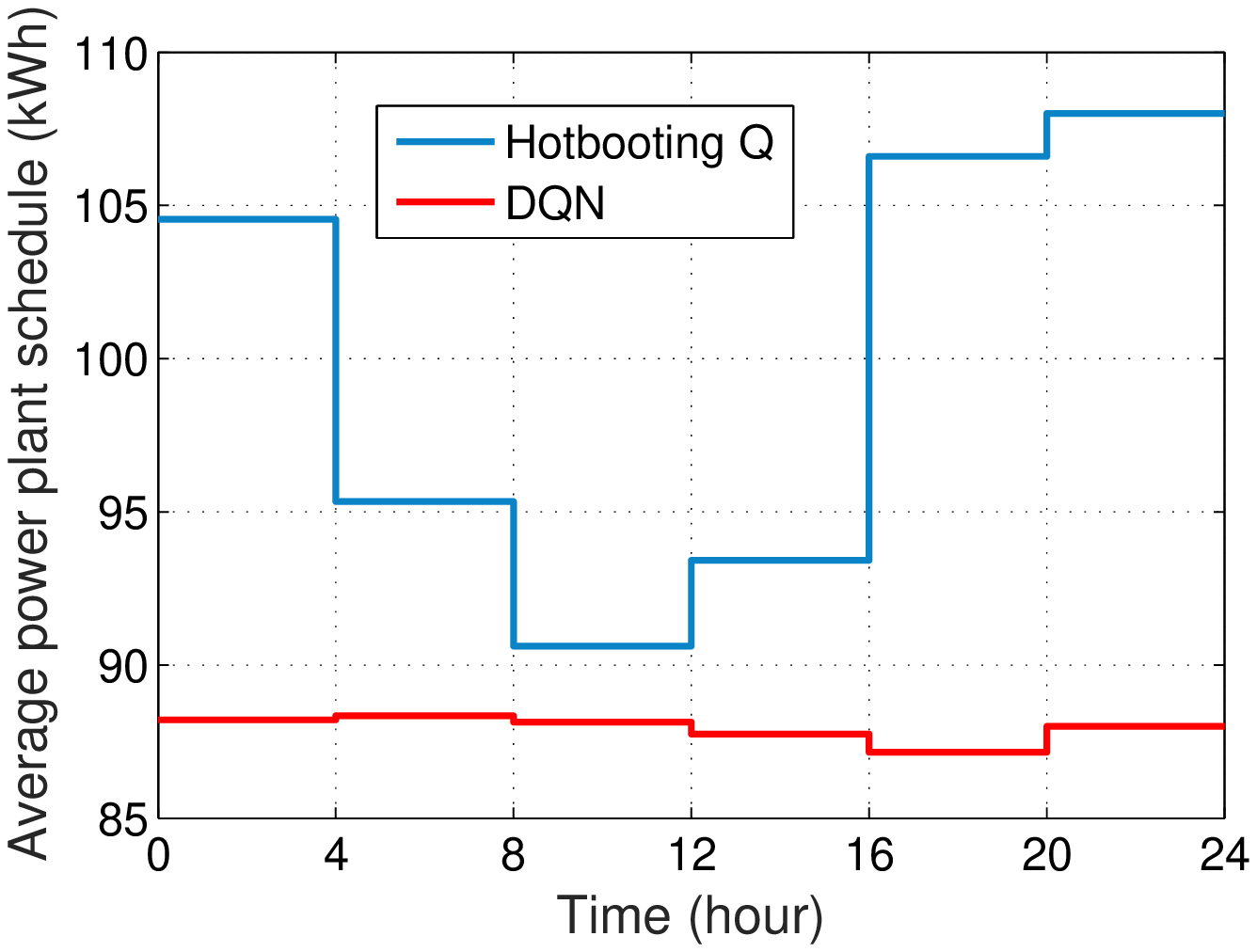}}
        \end{minipage}
    \begin{minipage}[t]{0.3\linewidth}
  \subfigure[Battery capacity, $B$]{
    \label{tradepp_cap} 
    \includegraphics[width=2.3in]{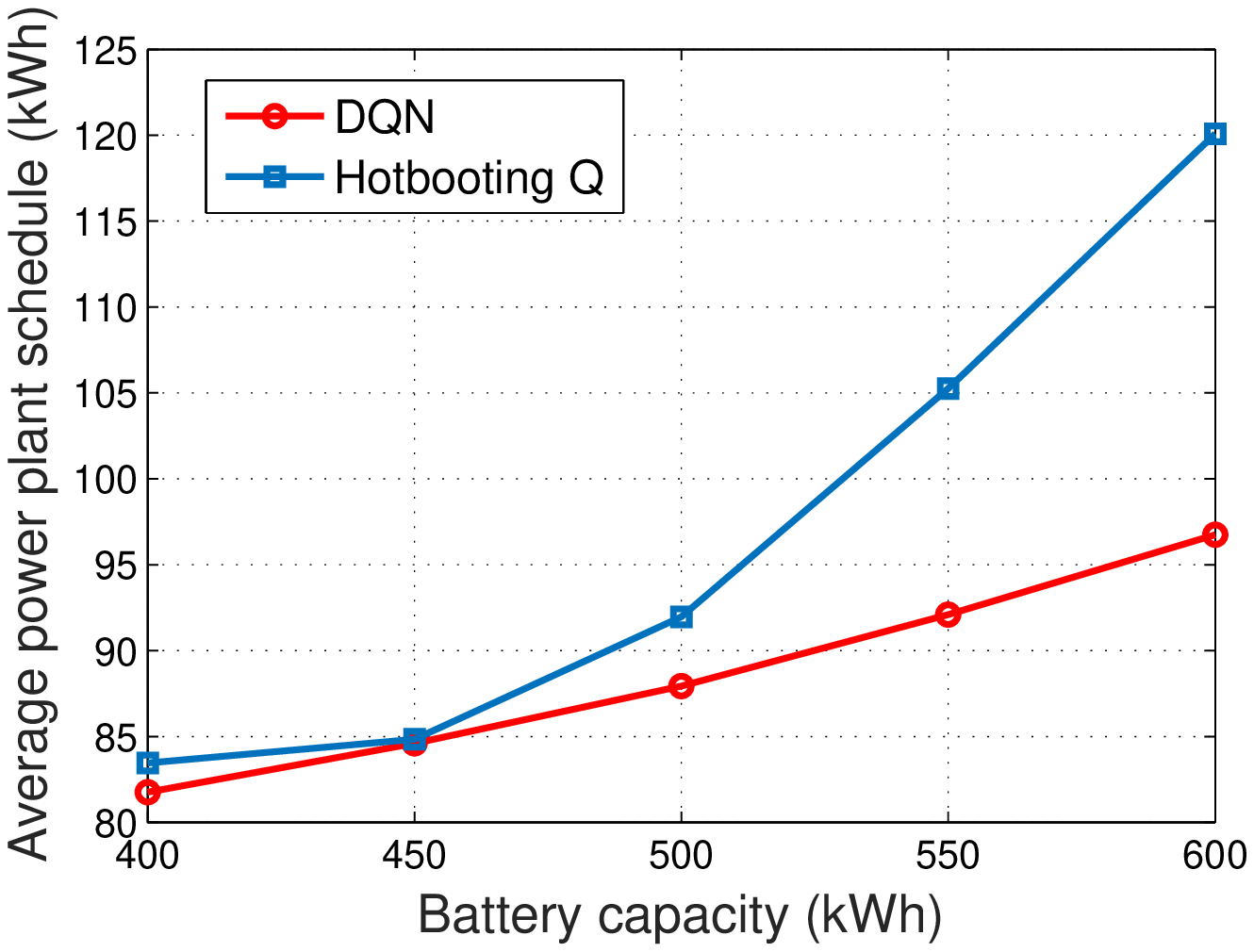}}
        \end{minipage}
    \begin{minipage}[t]{0.3\linewidth}
  \subfigure[Price ratio, $\varepsilon$]{
    \label{tradepp_price} 
    \includegraphics[width=2.3in]{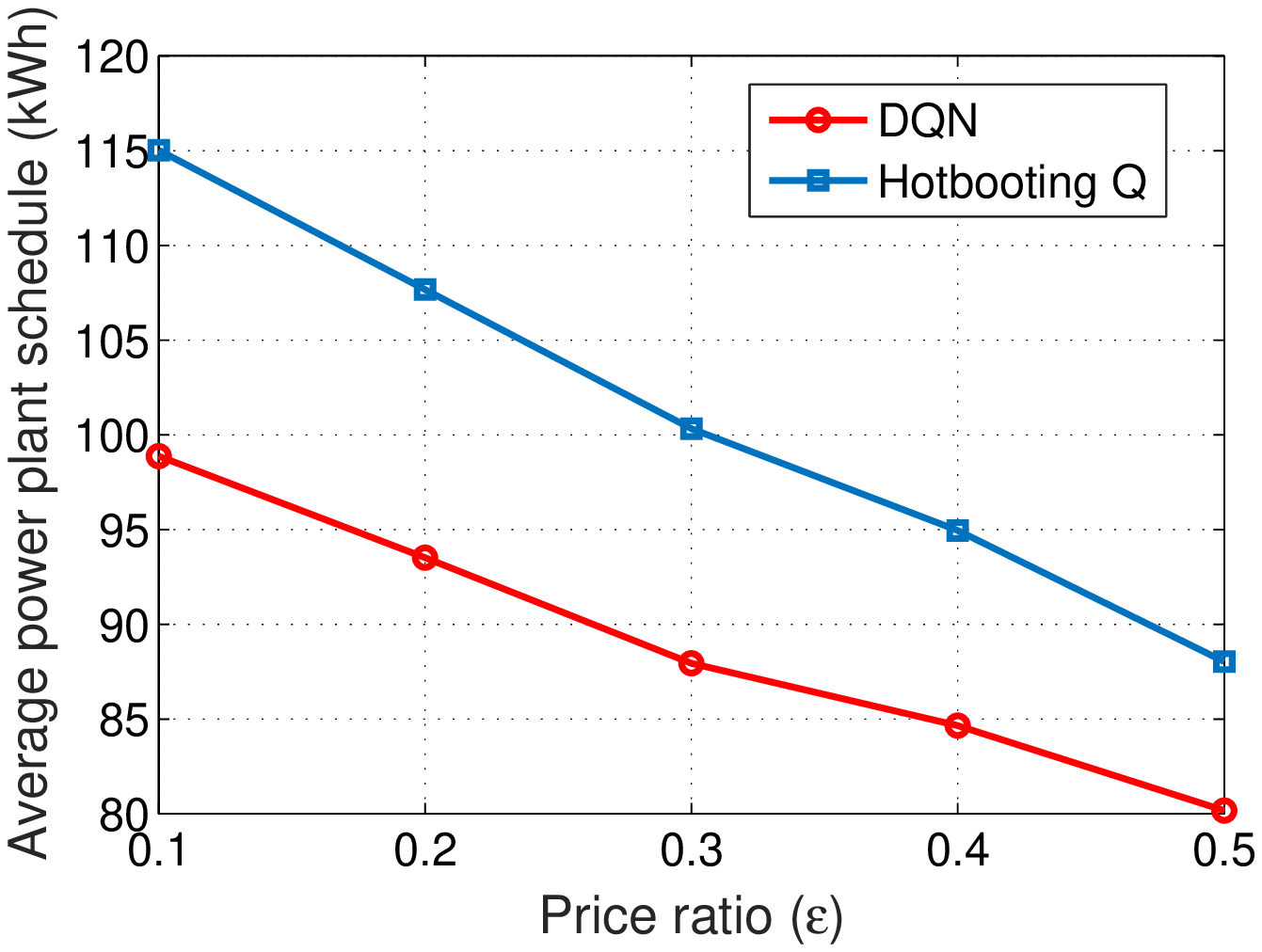}}
        \end{minipage}
  \caption{Average amount with the power plant.}
  \label{ave_energy} 
\end{figure*}

As shown in Fig. \ref{ave_energy}, the DQN-based energy trading scheme reduces the dependence on the power plant compared with  hotbooting Q. For example, the power plant schedule of the DQN-based strategy is 24\% lower at the night time (8:00 pm-12:00 pm), compared with the benchmark scheme.
The amount of the trading energy with the power plant decreases with the battery capacity, as MGs can save more energy at the low electricity price  and discharge the extra energy to support other MGs at the peak load time.
For example, this strategy decreases the average power plant schedule from 83 kWh to 96 kWh as the battery capacity increases from 400 to 600 kWh. The  power plant schedule with 600kWh capacity is 25\% lower than the  benchmark scheme, as shown in Fig. \ref{tradepp_cap}. As shown in  Fig. \ref{tradepp_price},  the power plant schedule decreases with the price ratio. For example, this strategy reduces the power plant schedule by 24\% as the price ratio increases from 0.1 to 0.5.

\begin{figure*}[htbp]
    \begin{minipage}[t]{0.3\linewidth}
  \centering
    \subfigure[Time in a day]{
    \label{utility_sim} 
    \includegraphics[width=2.3in]{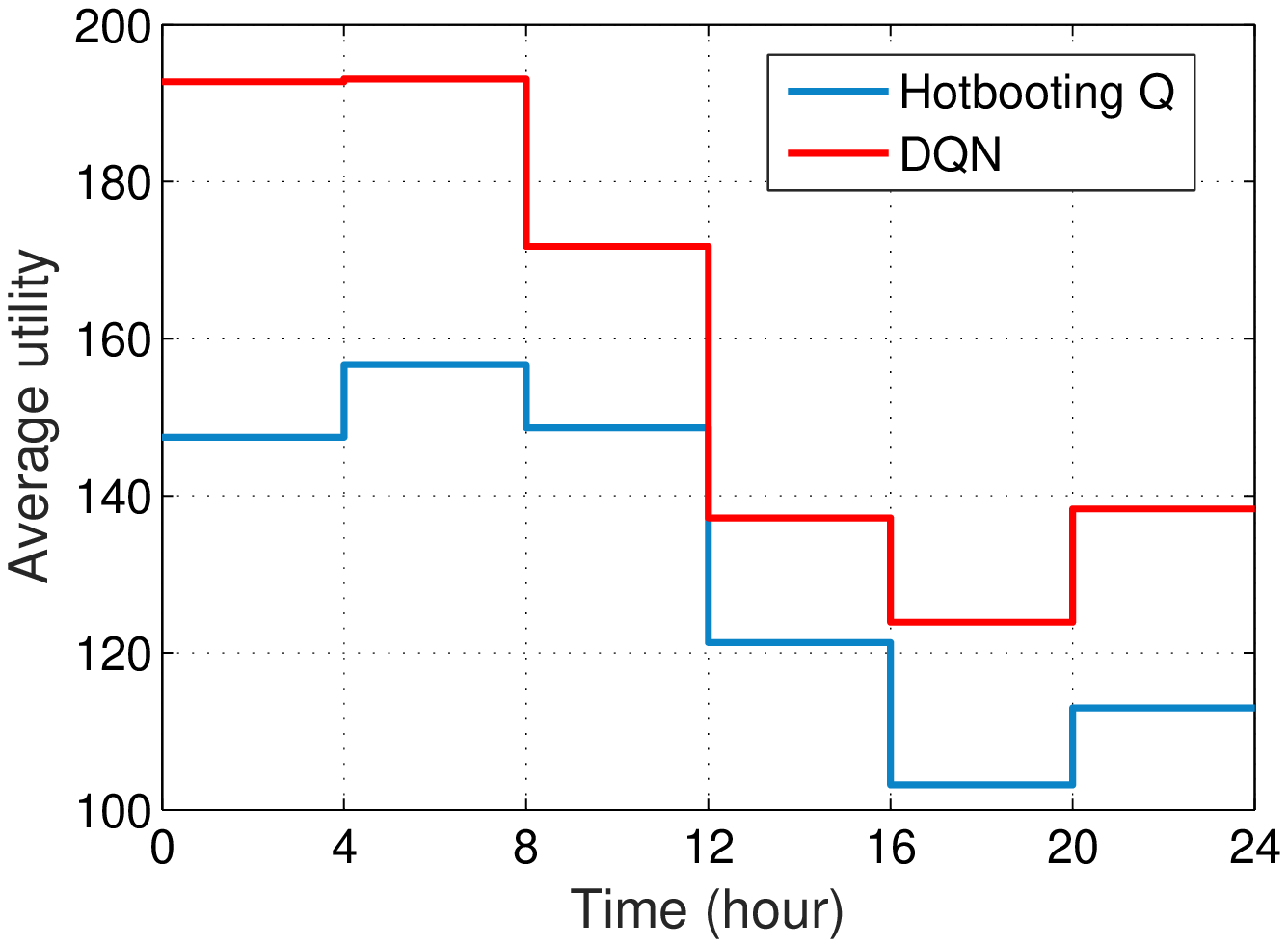}}
        \end{minipage}
    \begin{minipage}[t]{0.3\linewidth}
  \subfigure[Battery capacity, $B$]{
    \label{utility_cap} 
    \includegraphics[width=2.3in]{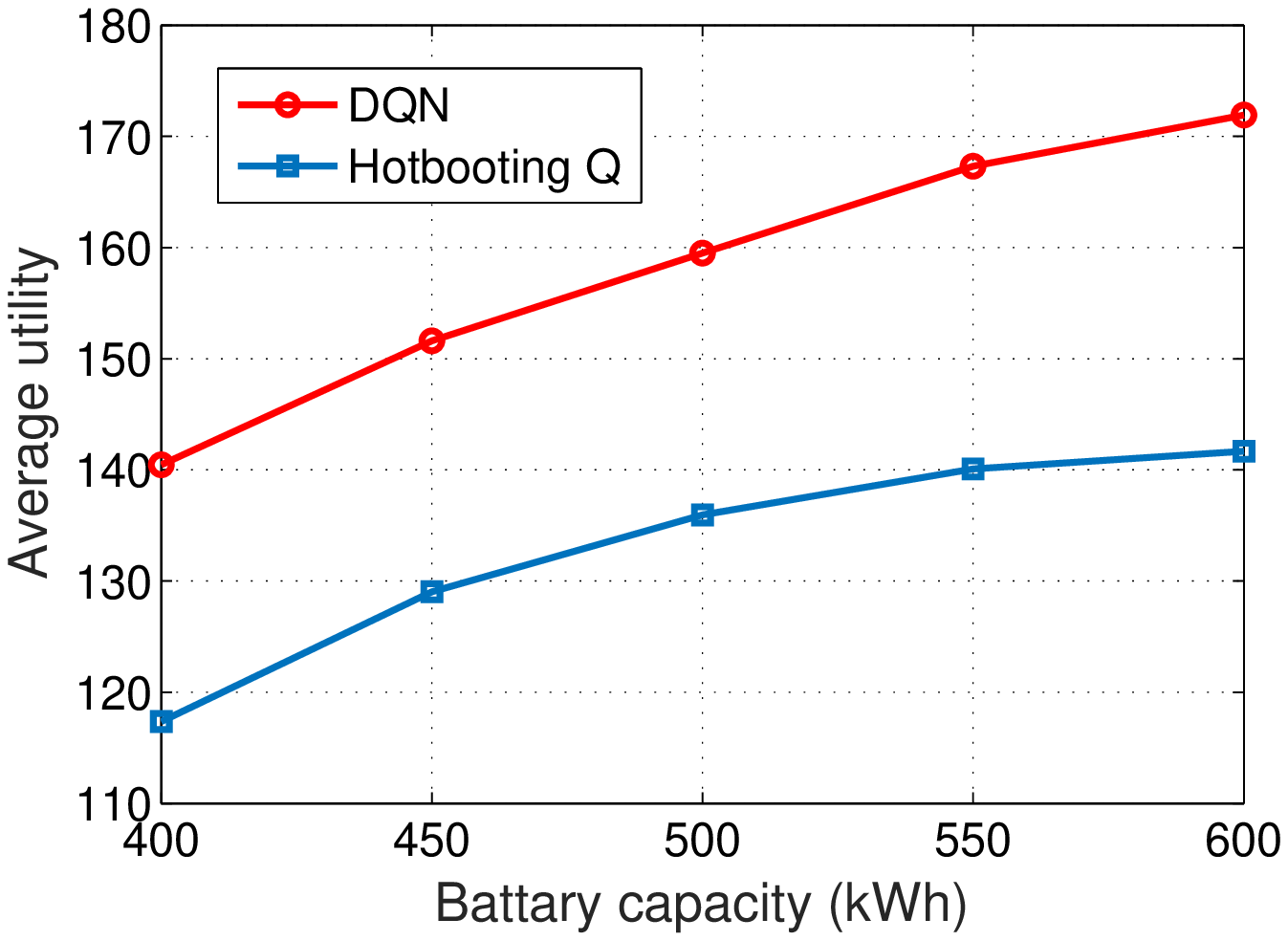}}
        \end{minipage}
    \begin{minipage}[t]{0.3\linewidth}
  \subfigure[Price ratio, $\varepsilon$]{
    \label{utility_price} 
    \includegraphics[width=2.3in]{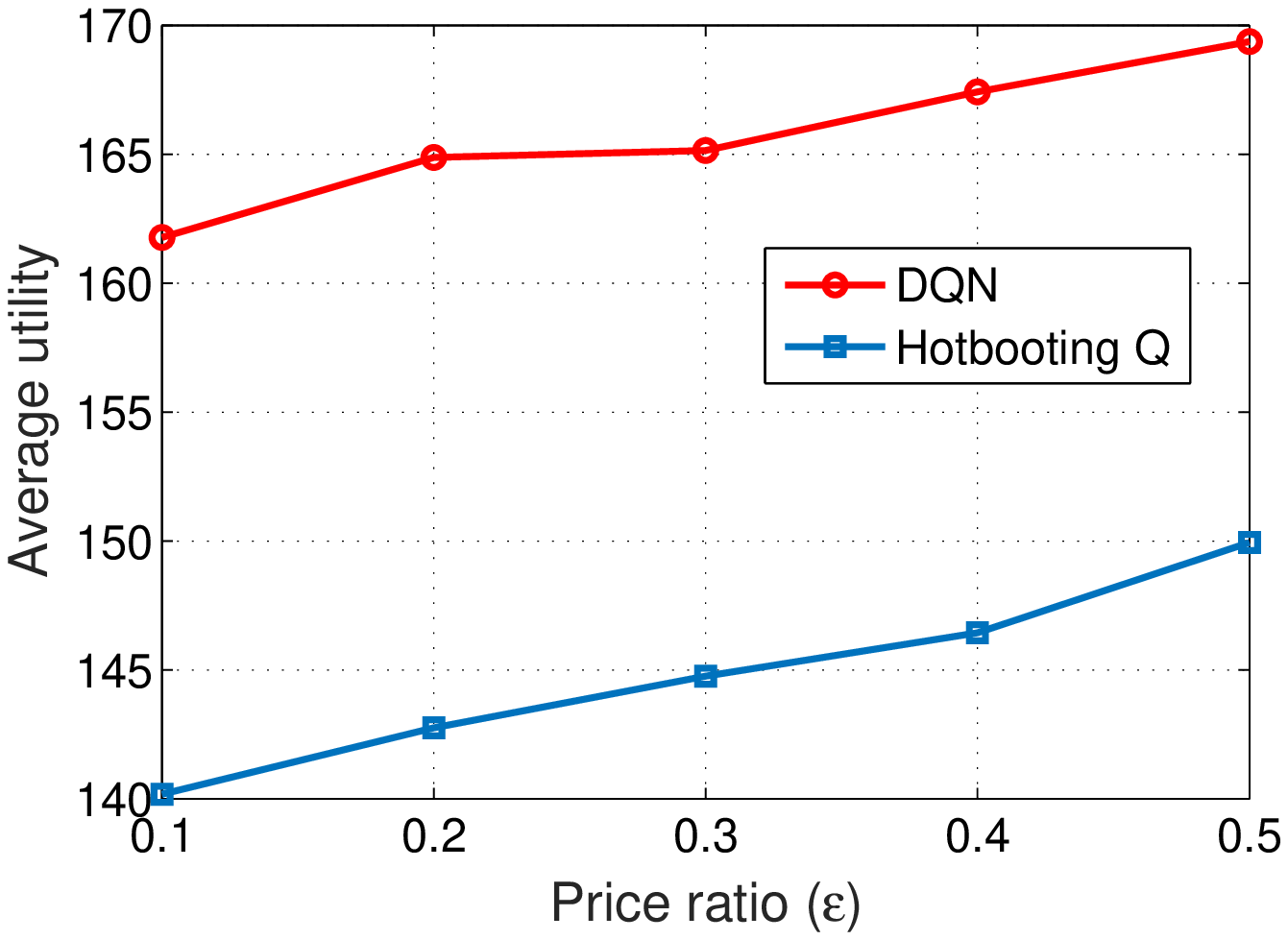}}
        \end{minipage}
  \caption{Average utility of an MG in a smart grid consisting of three MGs, with $\beta=120$.}
  \label{ave_utility} 
\end{figure*}

The DQN-based trading strategy improves the utilization of local renewable energy and increases the average utility of MGs, as shown in Fig. \ref{ave_utility}. For instance,  this scheme increases the average utility  by 29.7\% compared with the benchamark scheme during 00:00 am-04:00 am. If the battery capacity changes from 400 kWh to 600 kWh, the average utility of MGs increases by 22.8\%, which is 21.9\% higher than the benchmark
strategy. Finally, this strategy increases the average utility from 162 to 169 as the price ratio 0.1 to 0.5, and  the price ratio is 0.5, which is 13\% higher than that of the benchmark scheme.

\section{Conclusion}\label{conclusion}
In this paper, we have formulated an MG energy trading game for smart grids and provided the NE of the game, disclosing the conditions under which  a smart grid trade uses the renewable power generation of the MGs to satisfy the local power demand.
We have proposed a DQN-based energy trading strategy  to achieve the optimal energy trading policy in the dynamic game without aware of the energy generation and local demand models of the other MGs.  Simulations  based on realistic power generation and demand data demonstrate the effectiveness of this scheme, showing that this scheme can reduce the power plant schedule by 12.7\% and improve the utility of MGs by 22.3\%, compared with the benchmark strategy.

\appendices
%
%
%
%
%

\ifCLASSOPTIONcaptionsoff
  \newpage
\fi



%

\makeatletter
\newcommand{\adjustmybblparameters}{\setlength{\itemsep}{0\baselineskip}\setlength{\parsep}{0pt}}
\let\ORIGINALlatex@openbib@code=\@openbib@code
\renewcommand{\@openbib@code}{\ORIGINALlatex@openbib@code\adjustmybblparameters}
\makeatother

\bibliography{PU}
\bibliographystyle{IEEEtr}
\end{document}